\documentclass[preprint,12pt,authoryear]{elsarticle}

\usepackage{amssymb}
\usepackage{amsmath}

\usepackage{natbib}
\usepackage[colorlinks=true,linkcolor=black,citecolor=black,urlcolor=black]{hyperref}

\journal{Nuclear Physics B}

\begin{document}

\begin{frontmatter}


\author{Xavier Rodrigues}
\ead{rodrigues@apc.in2p3.fr}
\ead[url]{https://xrod.github.io}
\affiliation{organization={Université Paris Cité, CNRS, Astroparticule et Cosmologie}, 
           city={F-75013 Paris},
           country={France}}

\title{Neutrinos from extreme astrophysical sources}




\begin{abstract}
In this paper I review recent results on high-energy neutrino astronomy and what they can reveal about some of the most extreme cosmic accelerators. I discuss recent measurements of the diffuse TeV-PeV cosmic neutrino spectrum by the IceCube observatory and the current flux limits in the ultra-high-energy regime, contextualizing the recent detection of an ultra-high-energy neutrino by the KM3NeT observatory. I review the recent emergence of a TeV signal from nearby Seyfert galaxies such as NGC 1068, the potential of $\gamma$-ray blazars as neutrino sources above the PeV regime, and the current status of tidal disruption events and other transient classes as possible neutrino sources. For each of these topics, I discuss ongoing developments in source models and their current limitations. I argue for the indispensable role of next-generation multi-messenger facilities, such as IceCube-Gen2, in solidifying current source associations, probing the ultra-high-energy regime, and resolving vast transient populations that remain unidentified with current statistics.
\end{abstract}



\begin{keyword}
Astrophysical neutrinos \sep high-energy astrophysical transients \sep IceCube \sep KM3NeT \sep IceCube-Gen2 \sep active galactic nuclei \sep gamma-ray bursts \sep blazars \sep tidal disruption events \sep Seyfert galaxies



\end{keyword}

\end{frontmatter}

\section{Neutrinos as cosmic messengers}
\label{sec:intro1}

After six decades of observations of ultra-high-energy (UHE, >$10^{18}$~eV) cosmic rays, their sources remain unknown. The direct detection of UHE cosmic rays is limited to the local Universe, as they are deflected by magnetic fields and attenuated by interactions with the comic microwave background (CMB) and the extragalactic background light (EBL). To probe the deeper non-thermal Universe, we can take advantage of high-energy neutrinos, emitted in the course of hadronic interactions between cosmic-ray nuclei and ambient radiation \citep[cf.][]{Kelner:2006tc},
\begin{align}
p\,\gamma \,\rightarrow\,
\left\{
\begin{array}{l}
n\,\pi^{+} \;\rightarrow\; p\,e^+ e^-\,\nu_e\bar\nu_e\nu_\mu\bar\nu_\mu \\
p\,\pi^{0\,} \;\rightarrow\; p\,\gamma\gamma
\end{array}
\right.
\label{eq:pgamma_reaction}
\end{align} 
(and other non-resonant channels including multipion production at higher center-of-mass energies), or ambient gas,
\begin{align}
pp \,\rightarrow\,
\left\{
\begin{array}{l}
pn\,\pi^{+}\;(\,+\,\pi^0\,+\,\pi^+\pi^-\,+\,\cdots\,)\rightarrow\,\cdots\\
pp\,\pi^{0}\;\,\,(\,+\,\pi^0\,+\,\pi^+\pi^-\,+\,\cdots\,)\rightarrow\,\cdots
\end{array}
\right..
\label{eq:pp_reaction}
\end{align}
Both reactions lead to the production of high-energy neutrinos from the hadronic $\pi\rightarrow\mu\rightarrow e$ decay chain and from neutron decay, and $\gamma$-rays from the decay of the neutral pion. 

While the $pp$ reaction is kinetically feasible even for mildly relativistic nuclei, the $p\gamma$ process of Eq.~(\ref{eq:pgamma_reaction}) requires that the target photon has an energy in the rest frame of the relativistic nucleus of at least the pion mass. Neglecting pitch-angle effects on the kinematics, we can approximate the interaction threshold to  
\begin{equation}
\epsilon_\gamma^\prime \,\gtrsim\, \frac{m_\pi m_p}{E^\prime_p} \,\approx\, \frac{0.1\,\mathrm{GeV}^2}{E_p^\prime}
\,=\, \left(\frac{E_p^\prime}{\mathrm{100\,\mathrm{PeV}}}\right)^{-1} \mathrm{eV},
\label{eq:pgamma_threshold}
\end{equation}
where the prime signs indicate the rest frame of the interaction medium. Thus, TeV protons can interact at threshold with hard X-ray target photons; PeV protons with soft X-rays; and EeV protons with 0.1~eV infrared photons. Additionally to the near-threshold resonant process of Eq.~(\ref{eq:pgamma_reaction}), the total cross section also has contributions from multipion production channels, further increasing the interaction efficiency at energies above the kinematic threshold \citep{Mucke:1999yb}.

In resonant (near-threshold) $p\gamma$ interactions, the proton loses about 20\% of its kinetic energy to the pions (or 50\% in the case of $pp$ interactions). Thus, each of the neutrinos carries about 5\% of the parent proton's energy, and each $\gamma$-ray about 10\%. For a source at redshift $z$, relativistically boosted in the observer's frame by a Doppler factor $\delta_D$, the emitted neutrinos are therefore observed with energy
\begin{equation}
\epsilon_\nu\approx\,50\,\left(\frac{1}{1+z}\right)\left(\frac{\delta_\mathrm{D}}{10}\right)\left(\frac{E^\prime_\mathrm{p}}{\mathrm{100~PeV}}\right)\,\mathrm{PeV}.
\label{eq:pgamma_energies}
\end{equation}
This reaction also emits two $\gamma$-rays with $E_\gamma=100\,\mathrm{PeV}$, which can trigger electromagnetic cascades at the source or during propagation, sustained by pair-production interactions with low-frequency radiation~\citep{Berezinsky:2016feh}. This effect smears the original electromagnetic signal from hadronic interactions over a broad frequency range down to the electron mass scale, in the hard X-rays, or even lower frequencies in the case of magnetized sources where synchrotron emission is substantial.

On the contrary, the emitted neutrinos travel unscathed, presenting an opportunity for the identification of cosmic-ray interaction sites, even in sources lying at cosmological distances.
As a global network of high-energy neutrino experiments steadily expands, we are gradually overcoming the experimental challenges and statistical limitations inherent to the neutrino detection process. At the same time, the lack of a clear electromagnetic signal accompanying neutrino emission, due to the above-mentioned cascades, poses a challenge to the unambiguous  identification of the neutrino sources, which remains a subtle and hotly debated issue.

\section{Neutrino observations approach the ultra-high energies}
\label{sec:intro2}

The short burst of $\lesssim30\,\mathrm{MeV}$ neutrinos detected in coincidence with supernova 1987A provided a rare window into the process of stellar collapse~\citep{Kamiokande-II:1987idp}. In 2013, the IceCube South Pole observatory detected the first astrophysical neutrinos with energies in excess of a TeV, and up to at least a few PeV \citep{IceCube:2013cdw,IceCube:2013low}. The sources of these neutrinos have shown to be numerous and their identification is still an open challenge.
More recently yet, the KM3NeT experiment in the Mediterranean Sea detected the first event with a likely energy above 100~PeV \citep{KM3NeT:2025npi}, inaugurating the era of UHE neutrino astronomy.

IceCube is a cubic-kilometer experiment installed deep in the Antarctic ice, in continuous operation for over a decade~\citep[for a technical overview, cf.][]{IceCube:2016zyt}. The detection principle leverages the fact that a high-energy neutrino crossing the Earth can scatter off a nucleus via the weak interaction, producing a lepton (an electron, a muon, or possibly a tau lepton) that carries a large fraction of the neutrino's original energy. Charged secondaries emit Cherenkov radiation, detected by an array of over 5000 optical modules comprising the experiment. From the detected signal, the energy and direction of the original neutrino can then be estimated using sophisticated reconstruction algorithms.

KM3NeT, the successor of the ANTARES experiment in the Mediterranean sea, leverages the same general detection principle, with the exception that the array of optical modules is installed in liquid water, rather than ice. ARCA, KM3NeT's detector dedicated to high-energy cosmic neutrinos, is in a phase of completion and already in operation \citep{KM3Net:2016zxf}; in its final configuration, the detector volume will reach a cubic kilometer, equivalent to that of IceCube. P-ONE, a smaller-sized in-water experiment off the Canadian coast, is currently under deployment~\citep{Bailly:2021dxn}. These and other planned experiments represent an emerging worldwide network that will greatly complement and expand IceCube's view of the neutrino sky.

\begin{figure}
\includegraphics[width=\textwidth]{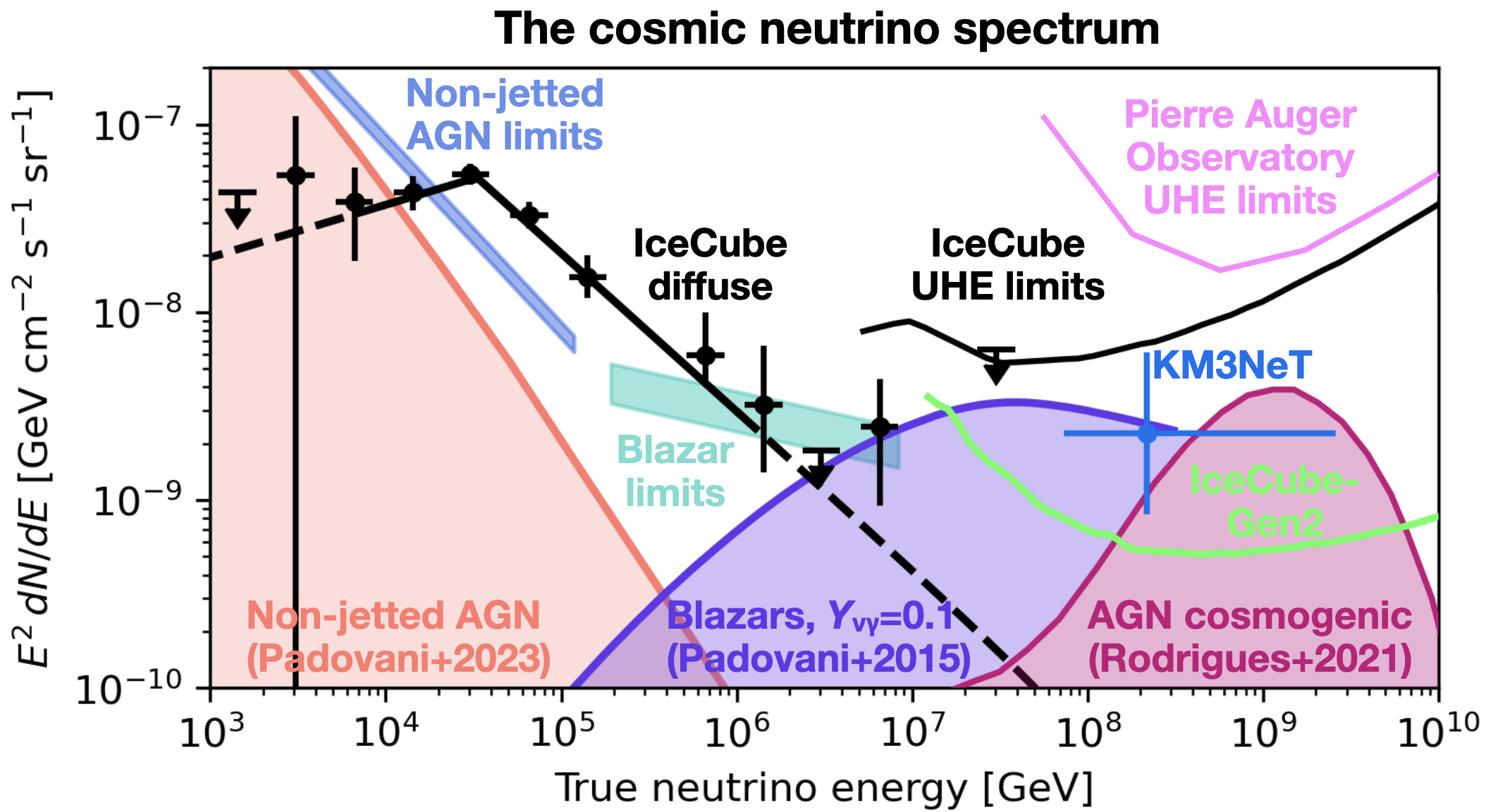}
\caption{Diffuse flux of all-flavor cosmic neutrinos between a TeV and a few PeV~\citep{IceCube:2025tgp} and current limits on the UHE flux by IceCube~\citep[][]{IceCubeCollaborationSS:2025jbi} and the Pierre Auger Observatory~\citep[][]{PierreAuger:2023pjg}. The blue data point is derived from KM3NeT's detection of the $>100$~PeV event KM3-230213A, taking into consideration the upper limits by other experiments~\citep{KM3NeT:2025ccp}. The three curves shaded underneath represent model predictions for the diffuse flux from X-ray-bright Seyferts~\citep[][]{Padovani:2024tgx}, $\gamma$-ray blazars~\citep{Padovani:2015mba} and cosmogenic neutrinos produced by UHE cosmic rays from a population of jetted AGN~\citep{Rodrigues:2020pli}. The model predictions are constrained by stacking limits on the population of X-ray-bright Seyferts \citep[blue band,][]{IceCube:2024dou} and $\gamma$-ray-bright blazars \citep[turquoise band,][]{IceCube:2016qvd}. The sensitivity of the future IceCube-Gen2 in the UHE range, enhanced by the planned in-ice radio array, is shown in bright green \citep{IceCube-Gen2:2020qha}.}
\label{fig:diffuse}
\end{figure}

The optimal energy window of these experiments lies between a few TeV and a few PeV. Below TeV, the astrophysical signal is drowned by the atmospheric backgrounds, i.e. the neutrinos and muons produced in the atmosphere by cosmic rays. Above a few PeV, the atmospheric background drops steeply, making the astrophysical signal background-free; but at those energies the number flux becomes challengingly low, even for cubic-kilometer experiments. The diffuse neutrino flux measured by IceCube within this energy window is represented in black in Fig. \ref{fig:diffuse}. As shown by a recent analysis, the cosmic neutrino spectrum in this range is well described by a broken power law, indicated here by the black lines~\citep{IceCube:2025tgp}. This has implications for the derived spectrum from extragalactic neutrino sources, and potentially eases previous tensions with the diffuse extragalactic  $\gamma$-ray spectrum when the neutrino spectrum was assumed to follow a single power-law~\citep{Murase:2015xka}. 
 
It is not yet clear what sources are responsible for the bulk of the diffuse IceCube flux. The $\sim$TeV signal recently identified by the IceCube collaboration along the Galactic plane \citep{IceCube:2023ame} shows that our Galaxy is a relatively poor neutrino emitter, consistent with the fact that the TeV-PeV cosmic neutrinos are of extragalactic origin. 
The recent IceCube detection of a point source signal from a few nearby Seyfert galaxies 
\citep{IceCube:2024dou} suggests that this source population may produce a considerable flux up to the TeV range, as exemplified by the salmon curve in Fig. \ref{fig:diffuse} \citep{Padovani:2024tgx}.  At the same time, as I discuss in Sec.~\ref{sec:seyferts}, those neutrinos are limited in energy to a few TeV, and the contribution of the X-ray-bright Seyfert population to the diffuse flux above these energies has been strongly constrained~\citep{IceCube:2024dou}, as shown in the figure by the blue band.

Above $\sim$100~TeV, the neutrino flux from the atmospheric backgrounds decreases dramatically~\citep{IceCube:2014rwe}, enabling a relatively high-purity astrophysical sample. Events at these energies also allow for the best angular reconstruction, with angular uncertainties $<1^\circ$ in the case of long tracks deposited by muons, and $<10^\circ$ in the case of cascades initiated by electrons. KM3NeT's ARCA promises to surpass this resolution by up to an order of magnitude, owing to the reduced photon scattering in liquid water compared to ice~\citep{KM3NeT:2024paj}.    

Upon detecting high-energy event candidates, both IceCube and KM3NeT send out real-time alerts to the multi-messenger community~\citep{IceCube:2016cqr,Assal:2021lbu}. In its first decade of operation, IceCube detected about 300 high-energy ``alert'' events~\citep{IceCube:2023agq}. 
A number of them has been determined to spatially overlap with $\gamma$-ray blazars, active galactic nuclei (AGN) with relativistic jets closely aligned with the line of sight. Blazars, and indeed jetted AGN at large, are theoretically expected to accelerate cosmic rays up to UHEs \citep[e.g.][]{Dermer:2008cy,Caprioli:2015zka,Blandford:2018iot}, and dedicated source models generally predict jetted blazars to be efficient emitters above the PeV range, as exemplified by the purple curve in Fig. \ref{fig:diffuse}.

In light of this theoretical expectation, the spatial overlap between blazars and high-energy IceCube events has prompted this source class as a major neutrino source candidate. However, the number of such events is prohibitively low, and no confident association with this source family has been established~\citep{IceCube:2023htm}. \citet{IceCube:2016qvd} have placed limits on the contribution of $\gamma$-ray blazars below  the PeV regime, as shown by the turquoise band in Fig. \ref{fig:diffuse}. As we can see, at about 200~TeV, the blazar contribution is limited to the $\sim10\%$ level.


The flux of TeV-PeV neutrino flux from other extreme source classes has also been constrained. These constraints are particularly severe in the case of bright transient phenomena for which the temporal window of expected neutrino emission is short. For example, for bright $\gamma$-ray bursts (GRBs), the most energetic cosmic explosions, IceCube could establish strong limits on their neutrino emission very early on, based on the lack of simultaneous detections \citep{IceCube:2012qza}. A neutrino follow-up of the binary neutron star merger event GW170817, the first gravitational wave event detected in association with a GRB \citep{LIGOScientific:2017ync}, revealed no simultaneous neutrino detection by IceCube, ANTARES, or the Pierre Auger Observatory~\citep{ANTARES:2017bia}. Today, the overall contribution of the GRB population to the diffuse flux of Fig.~\ref{fig:diffuse} is now known to be limited to at most the per-cent level~\citep{Abbasi:2022whi}. Given that GRBs are expected to accelerate protons or other nuclear species  \citep{Vietri:1995hs,Waxman:1997ti,Bottcher:1998qi,Murase:2011cx}, the non-detection of TeV-PeV neutrinos limits the efficiency of hadronic mechanisms at these energies. This, in turn, can translate into constraints on source models regarding the compactness of the dissipation region and the spectrum of the relativistic nuclei~\citep{He:2012tq,Bustamante:2016wpu,Liu:2022mqe}.

Altogether, these constraints imply that the bulk of the diffuse TeV-PeV cosmic neutrinos must originate in a vast cosmic population whose high-energy emission is either obscured, such as in the compact circum-nuclear regions of AGN~\citep{Stecker:2013fxa,Murase:2015xka,Murase:2019vdl,IceCube:2021pgw}, or intrinsically dim, such as transient events associated with low-luminosity GRBs~\cite[LLGRBs][]{Murase:2006mm,Liu:2012pf,Murase:2013ffa,Kashiyama:2013qet,Senno:2015tsn,Zhang:2017moz,Boncioli:2018lrv}, low-luminosity AGN~\cite[LLAGN][]{Kimura:2014jba,Murase:2014foa,Palladino:2018lov,Rodrigues:2020pli}, or low-luminosity tidal disruption events~\cite[TDEs,][see Sec. \ref{sec:tdes}]{Zhang:2017hom,Murase:2020lnu,Yoshida:2024fiu}. The limited statistics inherent to the neutrino detection process currently pose a critical challenge to the identification of such a vast number of sources. Ironically, this challenge stems from the same weakly interacting nature of neutrinos that makes them such appealing messengers from the distant cosmos.

On the other hand, above some 10~PeV the neutrino flux is still poorly constrained. \citet{IceCube:2025tgp} have derived the most constraining limits to date on the UHE diffuse neutrino flux, shown as a black curve in Fig.~\ref{fig:diffuse}. Above EeV, the Pierre Auger Observatory has a sensitivity comparable to that of IceCube, leading to flux limits at a similar level, as shown by the pink curve, adopted from \citet{PierreAuger:2023pjg} and rescaled to all-flavor neutrinos and 
decade-wide bins \citep[as in Fig. 1 of][]{IceCubeCollaborationSS:2025jbi}. Based on the IceCube limits available at the time, and assuming a pure-proton composition of the UHE cosmic rays, \citet{IceCube:2016uab} could constrain the cosmological evolution of the UHE cosmic-ray sources to values stronger than the star formation rate. Using the more recent limits shown in black in Fig. \ref{fig:diffuse}, \citet{IceCubeCollaborationSS:2025jbi} produced the first neutrino-based constraints on the UHE comic-ray mass composition, showing that the proton fraction above 30~EeV lies below 70\%, assuming the same strong cosmological source evolution.

At the same time, current limits still allow for a substantial flux of yet-to-be-identified UHE neutrinos (defined as above 100~PeV, in light of Eq.~(\ref{eq:pgamma_energies})). The recent KM3NeT detection of an UHE event suggests that UHE cosmic neutrinos may be within close reach; this is shown by the blue data point in  Fig. \ref{fig:diffuse}, derived from the KM3NeT detection of the single, >100~PeV event~\citep[and the non-detection of such events by other experiments, cf.][]{KM3NeT:2025ccp}. The origin of the event is not yet clear, but the KM3NeT collaboration has identified multiple candidate sources, including blazars, lying within the event's confinement region as currently estimated~\citep{KM3NeT:2025bxl}. Further detections will be essential to eventually test source models and further constrain the diffuse UHE neutrino flux.

This endeavor will benefit greatly from the planned array of new experiments targeting the UHE regime, the most ambitious of which is IceCube-Gen2, the next generation of the homonymous South Pole experiment~\citep{IceCube-Gen2:2020qha}. Aside from a ten-fold larger detection volume, which will considerably improve the sub-PeV sensitivity, Gen2 is also planned to include an array of radio antennas covering a vast area of the polar ice surface. By capturing the in-ice radio emission from charged secondaries, known as Askaryan radiation, this technique will deliver a dramatic improvement to our sensitivity in the UHE range,  shown  as a bright green curve in Fig. \ref{fig:diffuse}. Such sensitivity would provide an unprecedented probe of the cosmogenic neutrino flux, testing UHE source models and their cosmological evolution \citep[see as an example the  magenta curve in Fig. \ref{fig:diffuse}, showing a prediction for cosmogenic neutrinos from low-luminosity jetted AGN, cf. also][]{Kotera:2010yn,Murase:2015ndr,Heinze:2019jou,Rodrigues:2020pli}.

The first dedicated UHE neutrino experiment will likely be the Radio Neutrino Observatory in Greenland (RNO-G), already in deployment and planned to start operations around 2030~\citep{RNO-G:2020rmc}. Together with other future experiments, such as GRAND~\citep{GRAND:2018iaj} and the space-based POEMMA~\citep{POEMMA:2020ykm}, we may well have a network of UHE neutrino experiments around the globe by the end of the next decade. Given that UHE neutrino detection is generally constrained to zenith angles close to the experiment's horizon, such a global network will be a crucial asset to increase our field of view in this challenging energy range.

\section{Evidence mounts for TeV neutrinos from Seyfert galaxies}
\label{sec:seyferts}

In a recent breakthrough result, the IceCube collaboration reported an excess of $79^{+22}_{-20}$ neutrinos from a direction consistent with the source NGC~1068 \citep{IceCube:2022der}, one of the most nearby Seyfert galaxies. The region around the source displays the highest excess above background in the entire IceCube sky, and its spatial association with NCG~1068, at the 4.2$\sigma$ confidence level, is the most significant neutrino detection of an extragalactic source to date.

The neutrinos comprising the signal have energies between 1.5 and 15~TeV, implying, by Eq.~(\ref{eq:pgamma_energies}), a maximum energy of the primary protons of 30-300~TeV (assuming a relatively local source with no intrinsic Doppler boost). Given the absence of a bright radio jet, it has been proposed that these relativistic protons, or other nuclear species, may be accelerated in the corona, that is the hot, magnetized region lying close to the supermassive black hole, either by shocks \citep{Inoue:2019yfs} or in regions of magnetic reconnection or turbulence~\citep{Murase:2019vdl,Fiorillo:2023dts,Fiorillo:2024akm,Das:2024vug,Karavola:2024uui}. Shock acceleration in the larger-scale winds driven by the accretion system has also been suggested to play a role~\citep{Inoue:2022yak}.

\begin{figure}
\includegraphics[width=\textwidth]{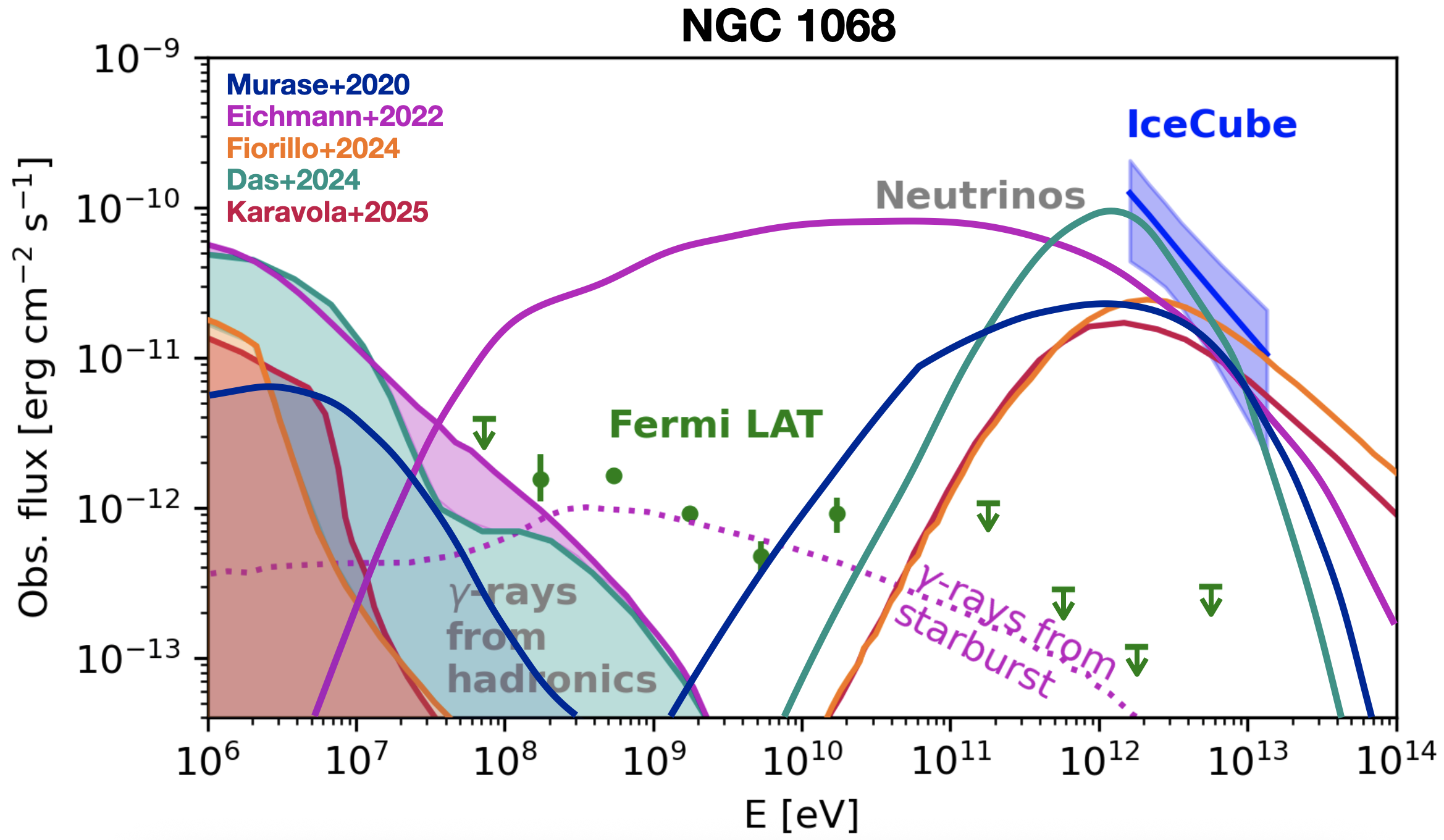}
\caption{Measurements of NGC~1068 in GeV $\gamma$-rays by the Fermi LAT (green data points) and best-fit neutrino flux by IceCube (blue band). The colored curves represent the neutrino spectrum (right-hand side) and the accompanying $\gamma$-ray spectrum (shaded underneath) predicted by different hadronic source models: \citet[][navy]{Murase:2019vdl}, \citet[][magenta]{Eichmann:2022lxh}, \citet[][orange]{Fiorillo:2023dts}, \citet[][teal]{Das:2024vug}, and \citet[][ruby]{Karavola:2024uui}. Since neutrinos should originate in an optically thick region, a less compact region must be responsible for the GeV $\gamma$-ray emission, in the case shown as a dotted curve, the starburst region surrounding the AGN core~\citep{Eichmann:2022lxh}.}
\label{fig:1068}
\end{figure}

The dense X-ray fields present in the corona, likely originating in the Comptonization of the thermal emission from the accretion system by hot electrons~\citep{Reis:2013tva,Fabian:2015wua}, make the environment optically thick to both the photoproduction of TeV neutrinos and the attenuation of GeV $\gamma$-rays via pair production. Both requirements are robust: the former to ensure energetic viability \cite[cf. the recent review by][]{Padovani:2024ibi}; the latter to describe the source's low $\gamma$-ray flux \citep{Inoue:2019yfs,Murase:2019vdl,Kheirandish:2021wkm}.
In fact, as I show in Fig. \ref{fig:1068}, the TeV neutrino flux derived from the IceCube signal (blue band) exceeds the GeV $\gamma$-ray flux measured by the Fermi LAT (green data points) by at least an order of magnitude. 

As we can see in Fig. \ref{fig:1068}, models of the compact circum-nuclear region can describe well the dramatic attenuation of the hadronically emitted $\gamma$-rays that is necessary to describe the low LAT fluxes. Below some 100~MeV, the predicted cut-off frequency and spectral shape of the escaping photons depends on the compactness of the emission region and the spectrum of the X-ray photons (cf. Eq.~(\ref{eq:pgamma_threshold})) assumed as  interaction targets.

As we can see by the blue band in Fig. \ref{fig:1068}, IceCube data suggests a soft neutrino spectrum in the TeV range. Hadronic models ascribe this feature to an intrinsic cutoff in the proton distribution~\citep{Murase:2019vdl,Das:2024vug} or, in the case of magnetic reconnection models, a spectral break whose energy depends on the proton magnetization~\cite{Fiorillo:2024akm,Karavola:2024uui}. Below the TeV range, as I explained in Sec. \ref{sec:intro2}, the neutrino spectrum is poorly constrained, owing to the strong atmospheric background. If protons are stochastically accelerated in the strong circum-nuclear turbulence their intrinsic spectrum should be hard, leading to a hard neutrino spectrum, as predicted by most models. However, softer neutrino spectra, such as that shown in magenta~\citep[][]{Eichmann:2022lxh} cannot technically be excluded by the data. As an alternative to the hadronic framework, \citet{Hooper:2023ssc} point out that TeV-scale neutrinos may also be produced leptonically, via the emission and subsequent decay of muon pairs from the annihilation between $\gtrsim$TeV $\gamma$-rays and the coronal X-rays.

In either scenario, it is clear that TeV neutrino emission from these sources must originate in a compact and optically thick region. Therefore, the observed GeV $\gamma$-rays should originate in a larger volume, such as the surrounding starburst region \citep[see dotted curve by][]{Eichmann:2022lxh}, as is known in general to be the case for star-forming galaxies~\citep{Ajello:2020zna,Fermi-LAT:2012nqz}. From a phenomenological  standpoint, this implies that GeV $\gamma$-rays and neutrinos from Seyferts are effectively independent messengers. Future $\gamma$-ray experiments targeting the MeV range, currently a blind spot, are therefore vital to ultimately test the hadronic corona framework and other source model classes.

Motivated by the relation between X-ray target photons and TeV neutrino emission predicted by models, IceCube expanded its search to a catalog of X-ray-bright Seyferts. Neutrino excesses have been found in association with a few other objects, including NGC~4151, with a combined likelihood of 2.7$\sigma$ \citep{IceCube:2024dou}. We may thus be witnessing the emergence of the first cosmic population of TeV neutrino sources. Owing to the isotropic nature of their emission, it is natural that the first associations have been with nearby sources. Further observation time and larger facilities, such as IceCube-Gen2, will be necessary to gradually probe more distant sources and solidify the current associations.

\section{Blazars as sources of PeV neutrinos and beyond}
\label{sec:blazars}

As introduced in Sec. \ref{sec:intro2}, a few object classes have been proposed as potential candidate sources of neutrinos above PeV and into the UHE regime, including LLGRBs~\citep{Senno:2015tsn,Zhang:2017moz}, starburst galaxies or galaxy clusters~\citep{Loeb:2006tw,Murase:2013rfa}, and the jets or cores of radio galaxies and AGN at large~\citep[][]{Stecker:1991vm,Bhattacharjee:1999mup,Blandford:2018iot}, possibly in  connection with magnetically arrested disks~\citep{Kimura:2020srn} or TDEs \citep[][see also Sec. \ref{sec:tdes}]{Dai:2016gtz,Lunardini:2016xwi}. In this section, I focus on jetted AGN, with an emphasis on recent observations and the efforts by the theoretical community to interpret them.

In the magnetized inner jet of an AGN, particles can be magnetically accelerated in sites of magnetic reconnection \citep{Sironi:2015eoa} or turbulence \citep{Kowal:2012rv,Comisso:2018kuh}, or along shear discontinuities \citep{Rieger:2004jz}. In regions of more moderate magnetization, shocks may likely drive the acceleration, as long suggested by observations of radio and optical knots~\citep{1978MNRAS.184P..61R} and, more recently, by X-ray polarization measurements~\citep{Liodakis:2022jio}.

With their relativistic jets aligned with the line of sight, blazars are observationally the most extreme class of AGN. The Doppler effect boosts the energy of the emitted photons and neutrinos by a factor of $\delta_\mathrm{D}=\Gamma$, for an on-edge viewing angle $\theta_\mathrm{obs}=1/\Gamma$, or up to $\delta_\mathrm{D}=2\Gamma$ for $\theta_\mathrm{obs}=0$. With measured Lorentz factors of $\Gamma=10$-50 \citep{Lister:2009zp}, the emission can be Doppler-boosted in energy by a factor of up to 100, and in number flux by up to $\delta^3=10^5$. For individual sources, this enormous flux boost can easily compensate for the luminosity distance suppression. Consequently, in contrast to Seyferts, there is little reason to expect neutrino associations with blazar jets to emerge preferentially from the nearest objects.

In 2017, IceCube detected an alert event with >100~TeV and good angular reconstruction spatially associated with the blazar TXS~0506+056~\citep{IceCube:2018dnn}, during a half-year-long period of enhanced multi-wavelength activity. At a confidence level of $3\sigma$ (later increased to $3.5\sigma$ following an archival search), this was largely regarded as the first detection of an extragalactic neutrino source.

In over a decade of IceCube observations, dozens of alert events have been detected in spatial association with other blazars. However, IceCube has consistently reported no significant association with the overall population of $\gamma$-ray-bright blazars 
[\citet{IceCube:2016qvd,Abbasi:2024ewg,Bellenghi:2023yza,IceCube:2023htm}, see also \citet{Franckowiak:2020qrq}], although multiple independent studies have reported intriguing correlations between publicly available IceCube events and different blazar catalogs \citep[e.g.][]{Giommi:2020hbx,Plavin:2020emb,Buson:2022fyf,Hovatta:2020lor}.

For the most part, this uncertainty is owed to limited statistics: while in the TeV regime neutrino experiments are overwhelmed by the atmospheric background, above 100 TeV the main challenge lies in the intrinsically low detection rate. As I mentioned in Sec.~\ref{sec:intro2}, IceCube detects a mere 30 alert events per year~\citep[][]{IceCube:2023agq}. With a ``signalness'', or probability of being astrophysical, in the 30-50\% range, this translates to an annual rate of only 10-15 high-energy astrophysical neutrinos. Given that the Fermi LAT detects thousands of $\gamma$-ray blazars~\citep{Fermi-LAT:2022oww}, the resulting per-source statistics is necessarily low and multiple detections from a given source, which would be essential for a definitive association, are unlikely.

\begin{figure}
\includegraphics[width=\textwidth]{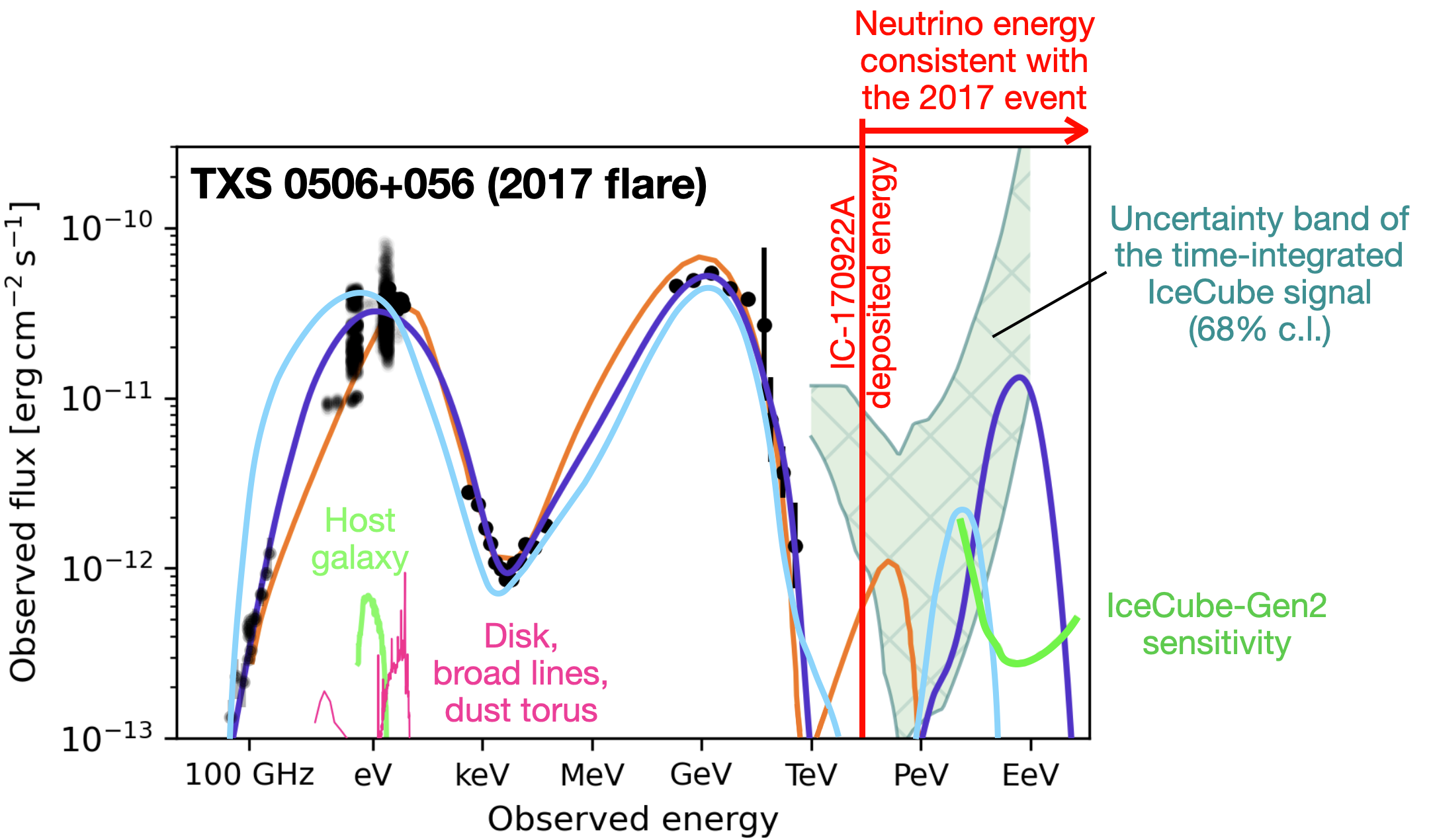}
\caption{Blazar TXS~0506+056 was the first source to be associated to a high-energy neutrino event in 2017, in temporal coincidence with a half-year-long multi-wavelength flare~\citep{IceCube:2018dnn}. The black data points show the quasi-simultaneous multi-wavelength measurements~\citep{IceCube:2018dnn,Keivani:2018rnh}. The colored curves show the multi-wavelength and neutrino predictions from three source models by \citet[][orange]{Gao:2018mnu}, \citet[][cyan]{Keivani:2018rnh}, and \citet[][purple]{Rodrigues:2025cpm}. As we can see, all three models can roughly describe the multi-wavelength emission, but predict neutrino energies differing by more than three orders of magnitude. The model predictions are consistent with the energy deposited by the 2017 event (vertical red line) and with the uncertainty limits on the point source signal, at the 68\% confidence level \citep{Rodrigues:2024fhu}, based on the publicly available ten-year point source data~\citep[][green band]{IceCube:2019cia}. The sensitivity of IceCube-Gen2 in the UHE range \citep{IceCube-Gen2:2020qha}, in light green, shows that future experiments will be crucial to test source models.}
\label{fig:txs}
\end{figure}

This limitation is compounded by degeneracies in source models. Phenomenological blazar models typically share two basic components: \textit{(1)} one or multiple dissipation regions along the relativistic jet where a spectrum of relativistic protons and electrons interact with synchrotron radiation, producing secondary particles through the processes introduced in Sec.~\ref{sec:intro1}; \textit{(2)} in some sources with a radiatively efficient accretion system, broad line emission from a circum-nuclear broad-line region and thermal emission from a dusty torus (pink in Fig. \ref{fig:txs}) may play a crucial role as additional targets for photohadronic interactions and pair production, enhancing neutrino emission and the development of electromagnetic cascades \citep[e.g.][]{Murase:2014foa,Keivani:2018rnh}.

Even within the above boundary conditions, models still have a great deal of freedom concerning the location of the particle acceleration and the maximum energy of the relativistic particles, because the mechanisms underlying acceleration and diffusion are poorly constrained, especially in distant sources. Breaking these degeneracies in the source physics is a monumental challenge, even with the wealth of existing multi-wavelength and multi-messenger data.

To illustrate this, I show in Fig. \ref{fig:txs} three multi-messenger predictions from different models of source TXS~0506+056. The relativistic protons accelerated in the jet reach maximum energies of 100~TeV, 10~PeV, and 100~PeV~\citep{Gao:2018mnu,Keivani:2018rnh,Rodrigues:2025cpm}. In all cases, the predicted neutrino spectrum is hard: this is a robust characteristic of the photohadronic blazar framework, driven by the resonant nature of photopion  production, the kinematic threshold of Eq. (\ref{eq:pgamma_threshold}), and the spectra of the available target photons. However, the maximum energy of the predicted neutrino spectrum varies among models by three orders of magnitude, from the sub-PeV range up to UHEs, since it follows the maximum energy of the primary protons. Strikingly, however, all three solutions predict broadband fluxes that are roughly compatible with multi-wavelength observations, including the simultaneous X-ray and $\gamma$-ray spectrum~\citep{IceCube:2018dnn}, making these models experimentally indistinguishable.

More remarkably yet, neutrino observations themselves cannot (yet) rule out any of the models. Like the vast majority of the IceCube alert events, the 2017 event associated with TXS~0506+056 was a throughgoing muon track, which means the original astrophysical neutrino interacted at a unknown location before reaching the detector. Thus, the 23.7$\pm2.8$~TeV deposited by the muon in the detector \citep[]
[]{IceCube:2018dnn} only implies a lower limit on the true energy of the original neutrino. All three models shown here predict a significant flux above $\sim100$~TeV, and are therefore compatible with the 2017 detection. 

The detection of high-energy starting events (i.e. neutrinos interacting inside the instrumented volume) associated with blazars would contribute greatly towards resolving this degeneracy, since in those cases the true neutrino energy can be well estimated. However, the effective area for high-energy starting events is drastically lower than for throughgoing tracks, making such associations unlikely~\citep[since the interaction volume is limited to the fiducial volume of the detector, cf.][]{IceCube:2020wum}. 

Additionally to individual high-energy track events, time-integrated IceCube point source analyses have also shown excesses associated with blazars, including with TXS~0506+056~\citep{IceCube:2019cia}. Intriguingly, the best fit of this time-integrated signal seems to favor <10~TeV neutrinos, a substantially different energy range compared to the $>100$~TeV event that triggered the source association in 2017. Given that leptohadronic source models generally predict hard spectra dominated by energies $\geq100$~TeV, as I described above, the emission of neutrinos below 100~TeV from blazars is currently an open challenge for theory.

To understand this challenge, it is important firstly to note that the emission of GeV $\gamma$-rays from blazars is typically well described by a dissipation region lying at a scales comparable or slightly larger than the broad-line region, some $10^3$ gravitational radii out from the supermassive back hole~\citep{Dermer:2008ug,Poutanen:2010he}. Here, the jet is sufficiently compact for the efficient production of GeV $\gamma$-rays and neutrinos, but not too compact to trap the $\gamma$-rays through pair production. The ambient radiation at these scales is dominated by hydrogen broad line emission~\citep{Greene:2005nj}, which appears in the rest frame of the jet with a Doppler-boosted energy of
\begin{equation}
{\epsilon_\gamma^{\mathrm{BLR}}}^\prime\approx100\,\left(\frac{\epsilon_{\mathrm{H\,Ly}\alpha}}{10.2\,\mathrm{eV}}\right)\left(\frac{\Gamma_\mathrm{jet}}{10}\right)\,\mathrm{eV},
\label{eq:blr}
\end{equation}
enabling, by Eqs. (\ref{eq:pgamma_threshold}) and (\ref{eq:pgamma_energies}), the efficient  emission of neutrinos with observed energy
\begin{equation}
E_\nu^\mathrm{BLR}\gtrsim 500\,\left(\frac{\delta_\mathrm{D}}{\Gamma_\mathrm{jet}}\right)\left(\frac{\epsilon_{\mathrm{H\,Ly}\alpha}}{10.2\,\mathrm{eV}}\right)^{-1}\,\mathrm{TeV}.
\label{eq:blr_neutrinos}
\end{equation}
Above these energies, neutrinos can also be produced at threshold via interactions with the infrared radiation from the dusty torus,
\begin{equation}
E_\nu^\mathrm{IR}\gtrsim 50\,\left(\frac{\delta_\mathrm{D}}{\Gamma_\mathrm{jet}}\right)\left(\frac{\epsilon_\mathrm{IR}}{0.1\,\mathrm{eV}}\right)^{-1}\,\mathrm{PeV},
\label{eq:blr_neutrinos}
\end{equation}
and at higher energies with the synchrotron photons present in the jet. On the contrary, neutrino emission considerably below the PeV range is severely limited by the scarce target photons~\citep[cf. also][]{Boettcher:2013wxa,Cerruti:2018tmc}. From the low levels of optical obscuration of blazar jets \citep[typical hydrogen column densities of order $10^{22}\,\mathrm{cm}^{-2}$ ][]{Willingale:2013tia}, we also know that the gas density in the jet is low, so hadronic interactions with ambient gas (Eq. (\ref{eq:pp_reaction})) are also inefficient. Significant $\lesssim100$~TeV neutrino emission from blazars would thus require large amounts of relativistic protons, with some models exceeding the source's Eddington limit by orders of magnitude~\citep[][orange curve in Fig. \ref{fig:txs}]{Gao:2018mnu}. Such a scenario is therefore disfavored from the point of view of source energetics.

It is also possible that a sub-PeV point-source signal from blazars has an entirely different origin from the high-energy photon emission. For instance, the jet may accelerate particles close to the base, in a compact and optically thick region, as in the Seyfert models discussed above. On the other hand, such a process would trigger efficient electromagnetic cascades, leading to strong X-ray emission. For a number of blazars, as in the example of Fig. \ref{fig:txs}, the characteristic X-ray ``spectral valley'' places tight constraints on such a scenario~\citep{Reimer:2018vvw,Rodrigues:2018tku}. It is therefore fair to say that if blazars are in fact responsible for neutrino emission both at tens of TeV and above PeV, there is currently no unified theoretical framework capable of describing it.

At the same time, the time-integrated point-source neutrino data are strongly dominated by lower-energy events from the atmospheric backgrounds, leading to large uncertainties. The excess spatially associated with TXS~0506+056, for instance, has a significance of $3.6\sigma$. As shown by the green band in Fig. \ref{fig:txs}, this uncertainty is not only in the strength, but also the energy, of the source signal, and at the 68\% confidence level, none of the three  models shown can be excluded ~\citep[cf.][for an technical discussion]{Rodrigues:2024fhu}. Further statistics are therefore crucial not only to confidently identify the associated sources, but to eventually construct an energy spectrum that can better constrain the source models.

\section{The rise and fall of three tidal disruption events}
\label{sec:tdes}

Another candidate class of TeV neutrinos that received considerable attention in recent years is that of TDEs, transients resulting from the gravitational disruption of a star transiting near a supermassive black hole. This phenomenon produces a flare in the UV, optical, and soft X-rays, with a sharp rise followed by a characteristic power-law decline \citep[cf.][for a recent review]{Gezari:2021bmb}. A small fraction of TDEs are also detected in radio, typically described by synchrotron emission from a jet or a sub-relativistic outflow~\citep{Alexander:2020xwb}. The viability of particle acceleration in shocks has long made TDEs promising candidates of cosmic rays and neutrinos~\citep{Wang:2015mmh,Dai:2016gtz,Senno:2016bso,Lunardini:2016xwi,Guepin:2017abw,Plotko:2024gop}.   

In 2019, IceCube detected a high-energy muon event in temporal coincidence with source AT2019dsg~\citep{Stein:2020xhk}, a TDE discovered by the ZTF survey and further detected in X-rays and in radio. The deposited energy suggested a neutrino of energy above $200\,\mathrm{TeV}$. Shortly after, two other high-energy events were detected in association with TDE candidates~\citep{Reusch:2021ztx,vanVelzen:2021zsm}, with a joint significance estimated at the $3.6\sigma$ level. 

Additionally to the temporal coincidence between the neutrino arrival and the optical flares, all three associated objects were noted to display a soft X-ray spectrum and low-luminosity radio emission~\citep{vanVelzen:2021zsm}, rare features among the general TDE population. This strengthened the case for the association, and subsequent theoretical works investigated the possible acceleration and interactions of nuclei in non-relativistic outflows~\citep{Murase:2020lnu} or in the jet~\citep{Winter:2020ptf}. The target photons for the leptohadronic interactions are provided by thermal emission in either the UV, optical, or X-ray band; following Eq.~(\ref{eq:pgamma_energies}), the predicted energy of the associated neutrinos varies between $\sim100$~TeV and the UHE range. As discussed in Sec.~\ref{sec:blazars}, this entire range is consistent with the detected high-energy track events.

The IceCube collaboration recently presented a new improved method for the spatial reconstruction of real-time alert events~\citep{IceCube:2025bst}, improving the precision of the reconstructed neutrino arrival direction by up to a factor of five. For a fraction of the events, the best fit of the arrival direction has also shifted considerably. For each of the three events previously associated with TDEs, the 90\% uncertainty contour is now incompatible with the source's position, thus disfavoring the associations~\citep{IceCube:2025uzh}. 

Naturally, the fact that the three TDE-neutrino associations are now disfavored cannot rule out neutrino emission from this source class. These detections have motivated theoretical works that suggested viable mechanisms of neutrino emission by TDEs, and only improved statistics can eventually shed light on their potential contribution to the cosmic neutrino flux. Upcoming optical surveys, such as the Vera Rubin Observatory~\citep{LSST:2008ijt}, together with dedicated real-time multi-messenger analysis pipelines~\citep{Nordin:2019kxt} will play a crucial role in increasing the sensitivity of future multi-messenger searches and eventually constraining neutrino emission from TDEs and other vast source populations.

\section{Conclusion}

The IceCube south Pole experiment has now measured cosmic neutrinos from TeV to a few PeV for more than 12 years, and the identification of their sources remains a major challenge. In this review I have discussed the multi-messenger potential of high-energy neutrinos as astrophysical messengers and highlighted some recent high-impact results. I focused on three emerging classes of extragalactic neutrino source candidates at different energies: Seyfert galaxies, jetted AGN, in particular of the blazar sub-class, and TDEs. The latter emerged as a promising neutrino source class following three spatial and temporal associations, but following a recent analysis, those associations are now disfavored.      

At the TeV scale, Seyfert galaxies are now major neutrino source candidates,  signals from NGC~1068 and a few other nearby objects begin to emerge in the time-integrated data. Models generally ascribe the emission to a compact region closely surrounding the supermassive black hole, where the $\gamma$-rays co-emitted with neutrinos in the pion production process are strongly attenuated above 100~MeV. Future observations in the MeV $\gamma$-ray range may provide a crucial complementary multi-messenger probe of the hadronic framework in these heavily obscured environments.

More extreme sources, such as blazar AGN, may be significant neutrino emitters above the PeV range, but a confident association remains elusive. Above 100~TeV, where neutrinos have good angular resolution and a high signal-to-background ratio, the astrophysical neutrino flux is low and only a few dozen high-energy events have been associated with blazars; in the all-energy time-integrated IceCube data, a potential contribution from jetted AGN has been limited to the $\sim$10\% level. For GRBs, the corresponding limits are stronger by an order of magnitude ($\sim$1\%), because their transient nature enables time-domain searches, resulting in more stringent constraints in the absence of associations. In spite of an apparently negative result, the non-detection in neutrinos of short transients such as GRBs can still place meaningful constraints on the source physics in some models, such as the  baryonic loading and the optical thickness to hadronic interactions. This is exemplified by the non-detection in neutrinos of GRB 170817A, the first GRB associated with a gravitational wave event in 2017.

It may appear odd that the most powerful known sources contribute so few of the observed cosmic neutrinos. However, it is important to keep in mind that the neutrinos produced in the photohadronic framework tend to closely follow the maximum energy of the parent nuclei. It is therefore natural to expect that neutrinos from UHE sources may, in fact, peak considerably above the PeV range, where current cubic-kilometer experiments have limited sensitivity. At sub-PeV energies, it is possible that UHE sources may emit a relatively low neutrino flux, easily overshone by a more numerous population of weaker sources, preventing a confident identification.

In spite of these challenges, the recent detection of a $>100$~PeV neutrino by the KM3NeT experiment provides a key indication that we may be approaching the discovery threshold for UHE neutrinos. The next generation of neutrino experiments planned worldwide will deliver a game-changing sensitivity in the UHE regime, and may likely push beyond that discovery threshold. The most extreme astrophysical sources are close at hand.

\section*{Acknowledgemensts}

I would like to thank Björn Eichmann, Damiano Fiorillo, Despina Karavola, Paolo Padovani, Walter Winter, and two anonymous reviewers for their comments on the manuscript. I acknowledge support by the investment program ‘France 2030’ launched by the French Government and implemented by the University Paris Cité as part of its program ‘Initiative d’excellence’ IdEx (ANR-18-IDEX-0001), which also funded the HERMES: multi-messengers of the Earth and the Universe project that contributed to this work.

\bibliography{bibliography}
\end{document}